\documentclass[a4paper,10pt,twoside]{cpc-hepnp}

\usepackage{multicol}
\usepackage{graphicx}
\usepackage{booktabs}
\usepackage{amssymb,bm,mathrsfs,bbm,amscd}
\usepackage[tbtags]{amsmath}
\usepackage{lastpage}

\begin{document}

\title{Conceptional Design of Heavy Ion Linac Injector for HIRFL-CSRm}

\author{%
\quad  Xiaohu Zhang$^{1,2;1)}$\email{zhangxiaohu@impcas.ac.cn}%
\quad Youjin Yuan$^{1}$
\quad Jiawen Xia$^{1}$
\quad Xuejun Yin$^{1}$ \\
\quad Dayu Yin$^{1}$
\quad Xiaoni Li$^{1}$
\quad Xiucui Xie$^{1}$
\quad Heng Du$^{1,2}$ \\
\quad zhongshan Li$^{1,2}$
}
\maketitle

\address{%

$^1$ Institute of Modern Physics, Chinese Academy of Sciences, Lanzhou 730000, China\\
$^2$ University of Chinese Academy of Sciences, Beijing 100049, China\\
}

\begin{abstract}
A room temperature heavy ion linac has been proposed as a new injector of CSRm (the main Cooler Storage Ring) at HIRFL (Heavy Ion Research Facility in Lanzhou), which is expected to improve the performance of HIRFL. The linac injector can supply heavy ion with maximum mass to charge ratio of 7 and injection kinetic energy of 7.272MeV/u for CSRm, and the pulsed beam intensity is 3emA with the duty factor of 3\%. Compared with the present cyclotron injector SFC (Sector Focusing Cyclotron), the beam current from linac can be improved by 10-100 times. As the pre-accelerator of the linac, the 108.48MHz 4-rod RFQ accelerates ion beam from 4keV/u to 300keV/u, which achieves the transmission efficiency of 95.3\% with 3.07m long vanes. The phase advance has been taken into account to analysis the error tolerance, and parametric resonance have been carefully avoided by adjusting the structure parameters. KONUS IH-DTLs, which follow the RFQ, accelerate the ions up to the energy of 7.272MeV/u and inject into HIRFL-CSRm. The resonance frequency is 108.48MHz for the first two cavities and 216.96MHz for the last 5 DTLs. The maximum accelerating gradient can reach 4.95MV/m in DTL section with the length of 17.066m, and the total pulsed RF power is 2.8MW. A new strategy, for the determination of resonance frequency, RFQ vane voltage and DTL effective accelerating voltage, is described in detail. The beam dynamics design of the linac will be present in this paper.
\end{abstract}

\begin{keyword}
heavy ion, linac injector, HIRFL-CSR, 4-rod RFQ, KONUS, IH-DTL
\end{keyword}

\begin{multicols}{2}
\section{Introduction}

The Heavy Ion Research Facility in Lanzhou (HIRFL) has been successfully upgraded with a multi-functional Cooler Storage Ring (CSR) at the end of 2007 [1]. As the only injector of HIRFL, Sector Focusing Cyclotron (SFC), which has served for nearly 60 years, can provide beam to the experiment terminal directly as well as act as the injector of Separated Sector Cyclotron (SSC) or CSRm. However, CSR has to be shutdown when SFC provides beam to other experiment terminal, which results in the low utilization of CSR. So it is essential to construct a new injector for CSR in the present facility.

In recent years, the linear accelerator has gradually taken the place of the cyclotron as the injector in the large scale scientific facility, such as GSI-UNLINAC [2], TRIUMF-ISAC [3] and RIKEN-RILAC [4]. Compared with the cyclotrons, the linacs have larger acceptance, higher transmission and higher accelerating gradient. So the heavy ion linac named CSR-LINAC is proposed as the injector of CSRm in HIRFL, shown in Fig.1. As can be seen, the SFC works for the SSC or one of the experiment terminals downstream at a time. Meanwhile the CSR-LINAC, as the full-time injector, can supply beam to CSR independently. Consequently, the double injector parallel mode (DIPM) can be achieved in HIRFL, which doubles the beam time of HIRFL. The new operation scheme is illustrated in Fig.2.
\end{multicols}
\begin{multicols}{2}
\begin{center}
\includegraphics[width=5cm]{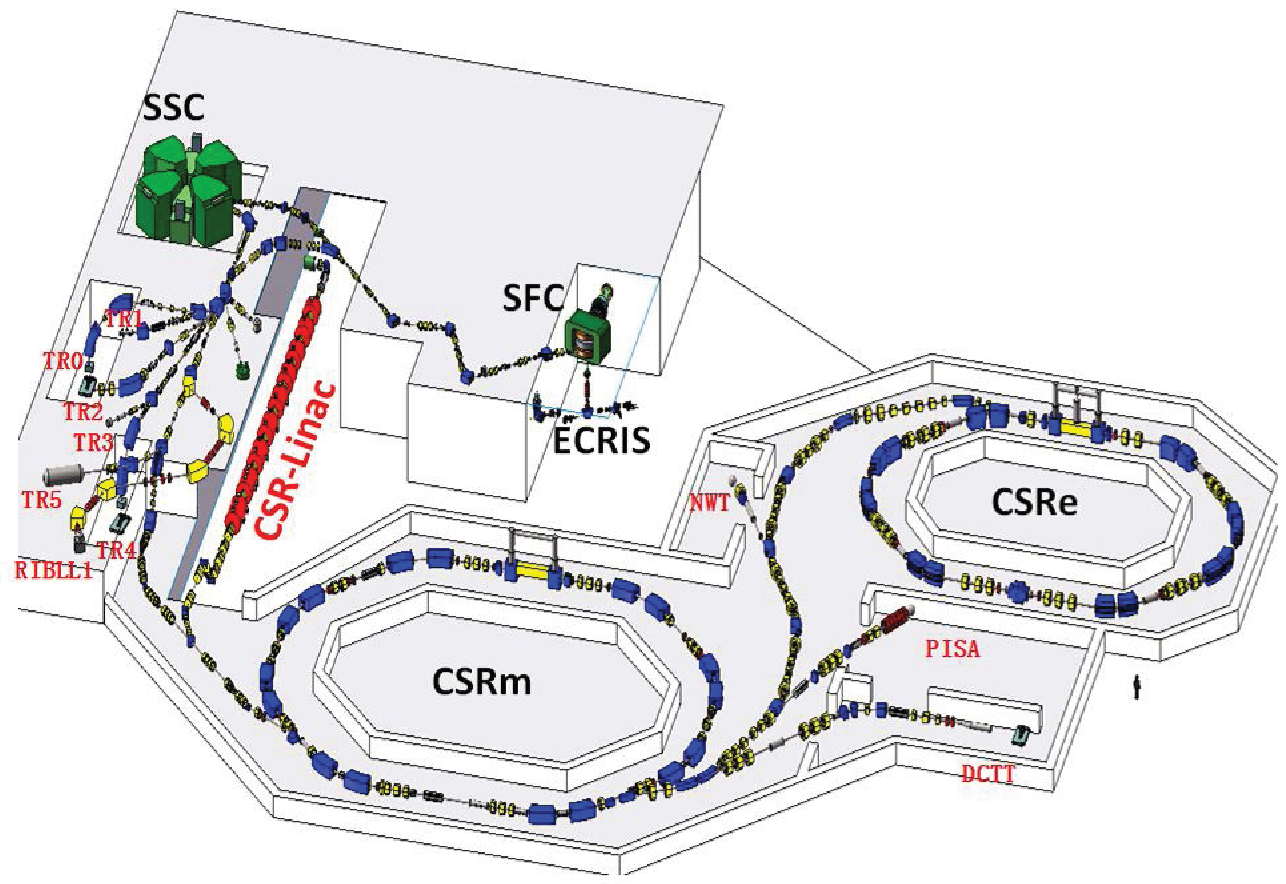}
\figcaption{\label{fig1} Layout of HIRFL-CSR with CSR-LINAC.}
\end{center}

 \begin{center}
\includegraphics[width=5cm]{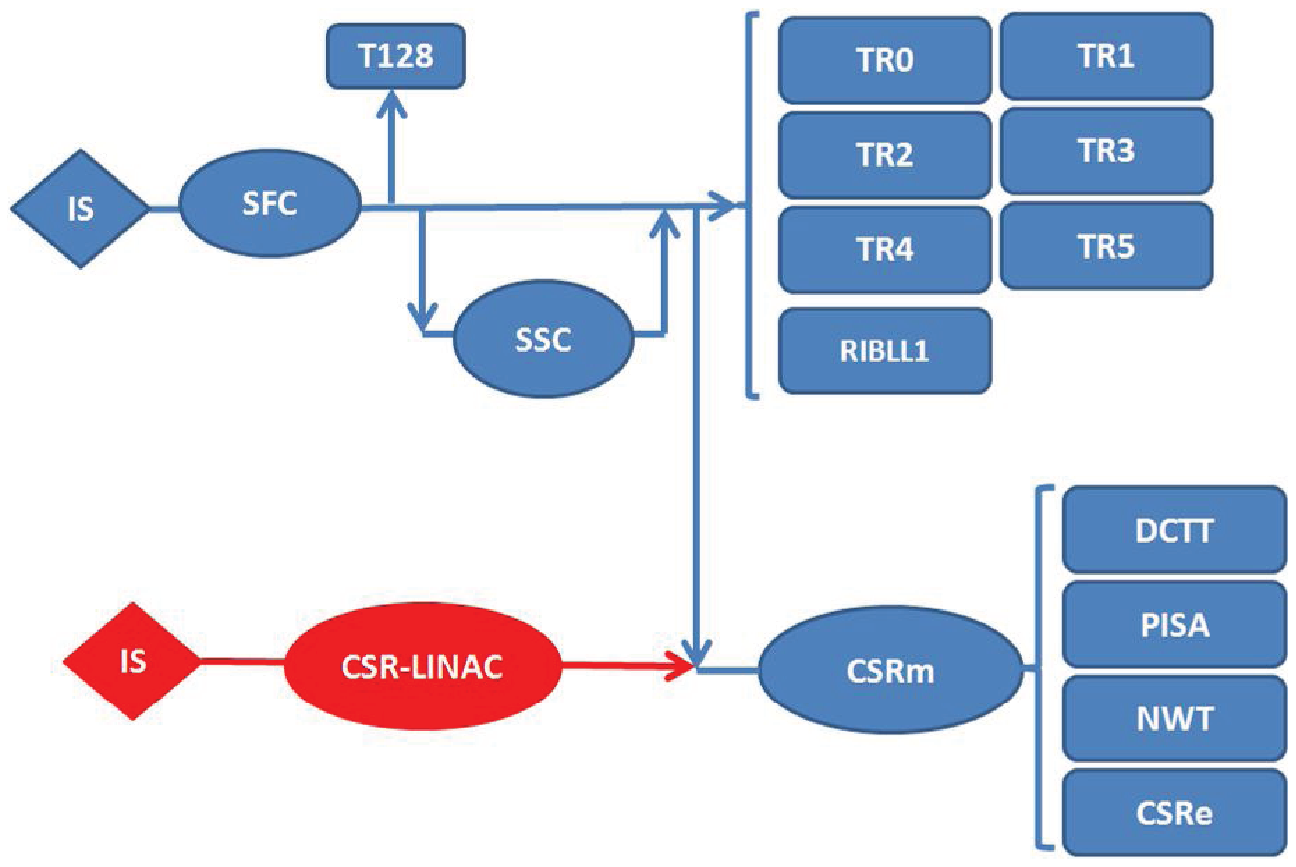}
\figcaption{\label{fig2} New operation scheme in HIRFL-CSR.}
\end{center}

\end{multicols}
\begin{center}
\includegraphics[width=15cm]{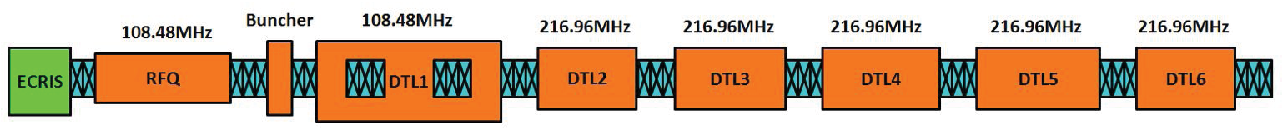}
\figcaption{\label{fig3} Layout of CSR-LINAC.}
\end{center}

\begin{multicols}{2}
CSR-LINAC accelerates all kinds of heavy ion from carbon to uranium at the resonance frequencies of 108.48MHz and 216.96MHz, and the pulsed beam current is 3emA with the duty factor of 3\%. The main parameters of CSR-LINAC are listed in table.1. Fig.3 shows the layout of the whole linac with the resonance frequencies for tanks respectively. The linac injector mainly includes the following components: (1) an Electron Cyclotron Resonance Ion Source(ECRIS) that generates 0.004MeV/u heavy ions; (2) a Radio-Frequency Quadrupole (RFQ) accelerator to bunch and pre-accelerate the ions to 0.3MeV/u; and (3) a Drift Tube linac (DTL), which is composed of 6 Interdigital H-type (IH) resonators, for the main acceleration up to 7.272MeV/u.

\begin{center}
\tabcaption{\label{tab1}  Main Parameters of the CSR-LINAC.}
\footnotesize
\begin{tabular*}{80mm}{c@{\extracolsep{\fill}}ccc}
\toprule
A/Q & 3 - 7 & - \\ \hline
Emittance(norm, 99\%)  & 0.88  & $\pi.mm.mrad$ \\ \hline
Frequency  & 108.48/216.96 & MHz \\ \hline
Pulsed beam current  & 3 & emA \\ \hline
Duration  & 3 & ms \\ \hline
Repetition & 10  & Hz \\ \hline
RFQ input/output energy & 0.004/0.3 & MeV/u \\ \hline
DTL input/output energy & 0.3/7.272 & MeV/u \\ \hline
Transmission(design) & $>$90 & \% \\
\bottomrule
\end{tabular*}
\end{center}

\section{The improvement of HIRFL-CSR}

Compared with SFC, CSR-LINAC has larger acceptance and higher transmission. According to the present performance, SFC can supply carbon with beam current of less than 10euA and only 0.68euA uranium beam for CSRm. The SFC transmission is so low because of small beam acceptance, low capture efficiency and uncertain orbit in the SFC. By contrast, the beam current of carbon ion from CSR-LINAC can reach nearly 150euA and uranium beam current approaches 60euA. Supposed that the multi-turn-injection scheme is adopted with electron cooling, the maximum beam intensity stored in CSRm is as 100 times as that at the injection. The maximum stored particle number comparison for SFC and CSR-LINAC as the injector of CSRm is shown in Fig.4. As can be seen, The maximum stored particle number of carbon ion has reached the space charge limit, $5.15\times10^{10}$, which is as nearly 11 times as that in the case of SFC as the injector of CSRm. Even for uranium, the maximum stored particle number can reach $4.85\times10^9$.

\begin{center}
\includegraphics[width=7cm,height=4cm]{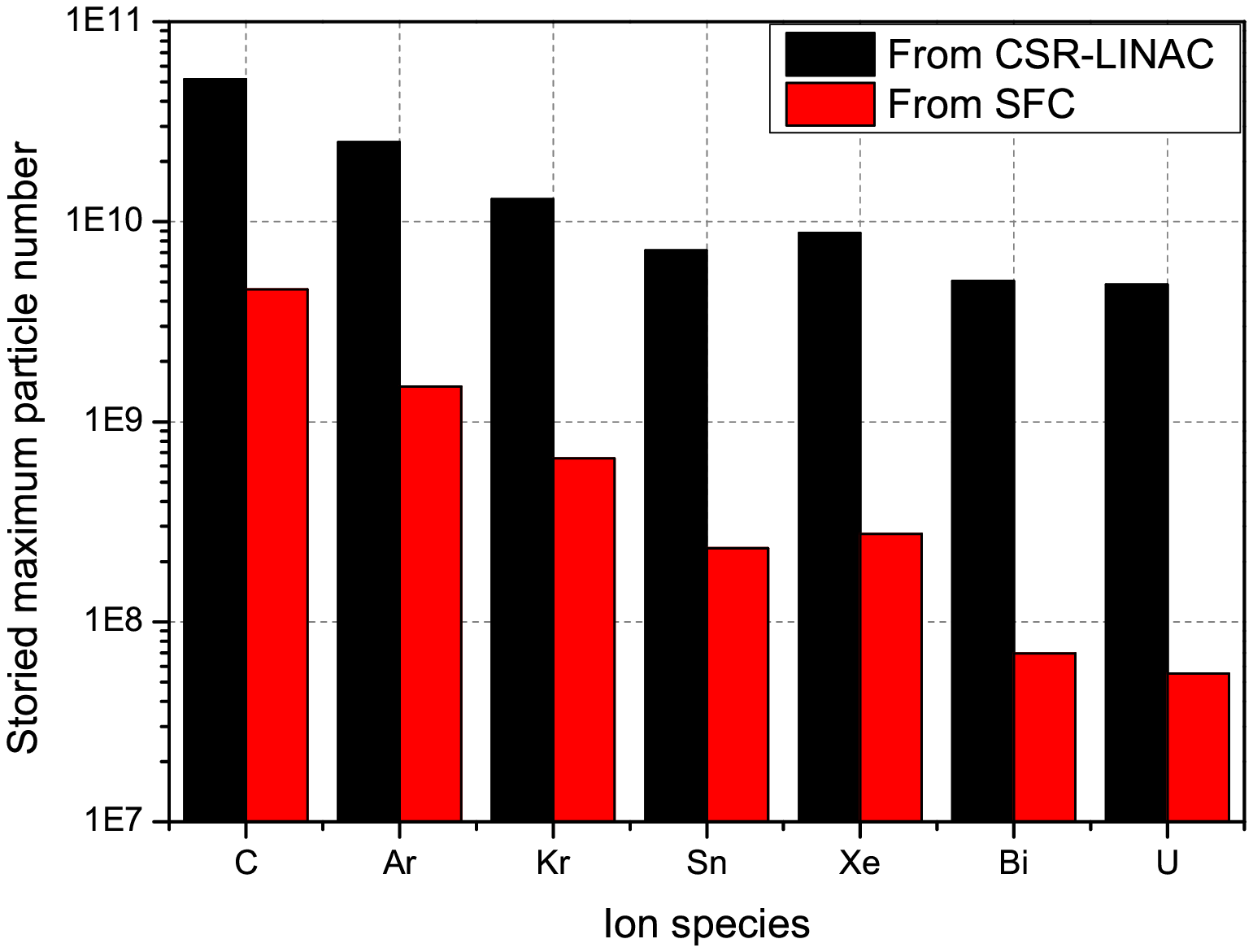}
\figcaption{\label{fig4} The maximum stored particle number comparison for SFC and CSR-LINAC as the injector of CSRm.}
\end{center}

The maximum magnetic rigidity is the characteristic parameter related to accelerating capability of synchrotron. The formula which calculates the magnetic rigidity is followed:
\begin{equation}
B\rho = \frac{A\sqrt{2E_rE_k+{E_k}^2}}{300Z}
\end{equation}
for the CSRm, the maximum magnetic rigidity is the constant value, 11.5 T.m. The final extraction energy can be improved by increasing the charge state of super-heavy ion. However, the heavier ion corresponds to the lower stripping efficiency [5]. Figure.5 shows the equilibrium charge state distribution of uranium ion with the energy of 7.272MeV/u. As can be seen, the stripping efficiency of ${}^{238}\!U^{34+}$ to ${}^{238}\!U^{67+}$ is only 18.27\% , which makes it essential to supply enough beam intensity before stripping. According to formula (1), the CSRm can accelerate stripped ${}^{238}\!U^{67+}$ to 414.22MeV/u but for ${}^{238}\!U^{34+}$ only to 122.3MeV/u. \\
\begin{center}
\includegraphics[width=7cm,height=4cm]{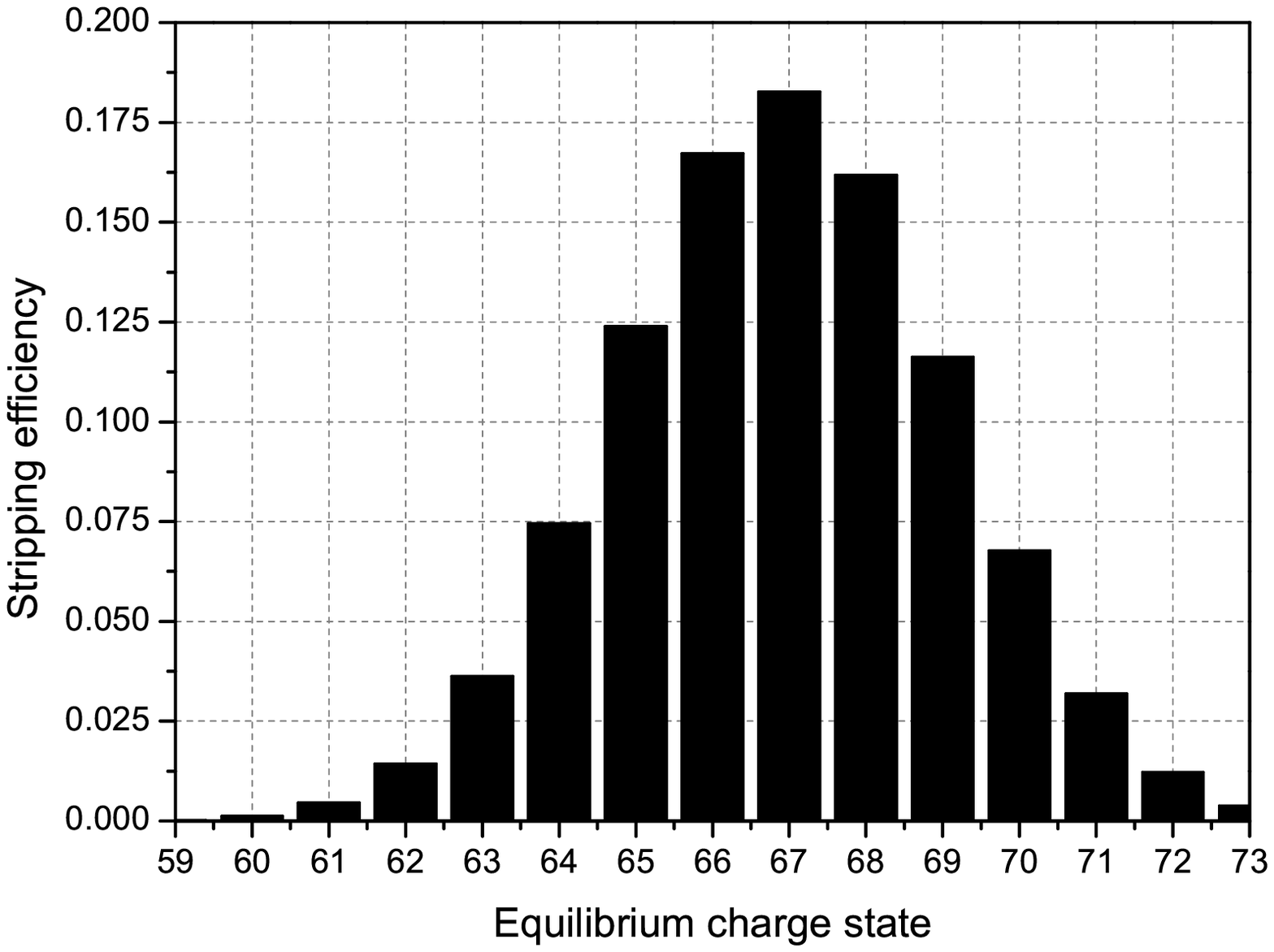}
\figcaption{\label{fig5} The equilibrium charge state distribution of ${}^{238}\!U^{34+}$ with the energy of 7.272MeV/u.}
\end{center}

\section{RFQ beam dynamics}

For a large linac complex, the RFQ accelerator at the front end has to change the continues input beam to suitable bunches with expected energy gain for the DTL accelerating structure downstream. According to the RFQ principle [6], the key parameter Br is introduced to determine a reasonable resonance frequency ($fr$):
\begin{equation}
Br = \frac{Ze{E_s}}{A{m_0}{f_r}^2\kappa}
\end{equation}
where B is the transverse focusing strength, r is the average aperture radius, $E_s$ is the maximum surface electric field on the pole and $\kappa$ is set to 1.36, which is reasonable in the RFQ structure. According to formula (2), Br is only proportional to Z/A and inversely proportional to square of ${f_r}$ for a given maximum surface electric field ($E_s$, 2 times the kilpatrick limit), illustrated in Fig.6. As can be seen, a Br value corresponds to a sole ${f_r}$ at a certain designed particle and therefore to choose a reasonable resonance frequency means to choose a reasonable Br. There are some considerations about Br as following:
\begin{itemize}
  \item a higher accelerating gradient at a higher resonance frequency, which is helpful to realize a compact RFQ structure, requires a smaller Br parameter;
  \item a larger Br parameter supplies a larger beam acceptance of RFQ channel, which means a better transmission performance;
  \item a larger Br parameter corresponds to freer selection of structure parameters.
\end{itemize}

Fig.6 shows different kinds of RFQs in international large scale scientific facilities. As can be seen, HSI-RFQ (14.6), HLI-RFQ (15.93) and HIT-RFQ (15.2) in German represent the lower limit value of Br ,which are nearly the highest level of RFQ design today. So Br of 20.15 is reasonable for heavy ion RFQ with minimum charge to mass of 1/7 in CSR-LINAC and corresponding 108.48MHz is set as the resonance frequency of RFQ.
\begin{center}
\includegraphics[width=7cm]{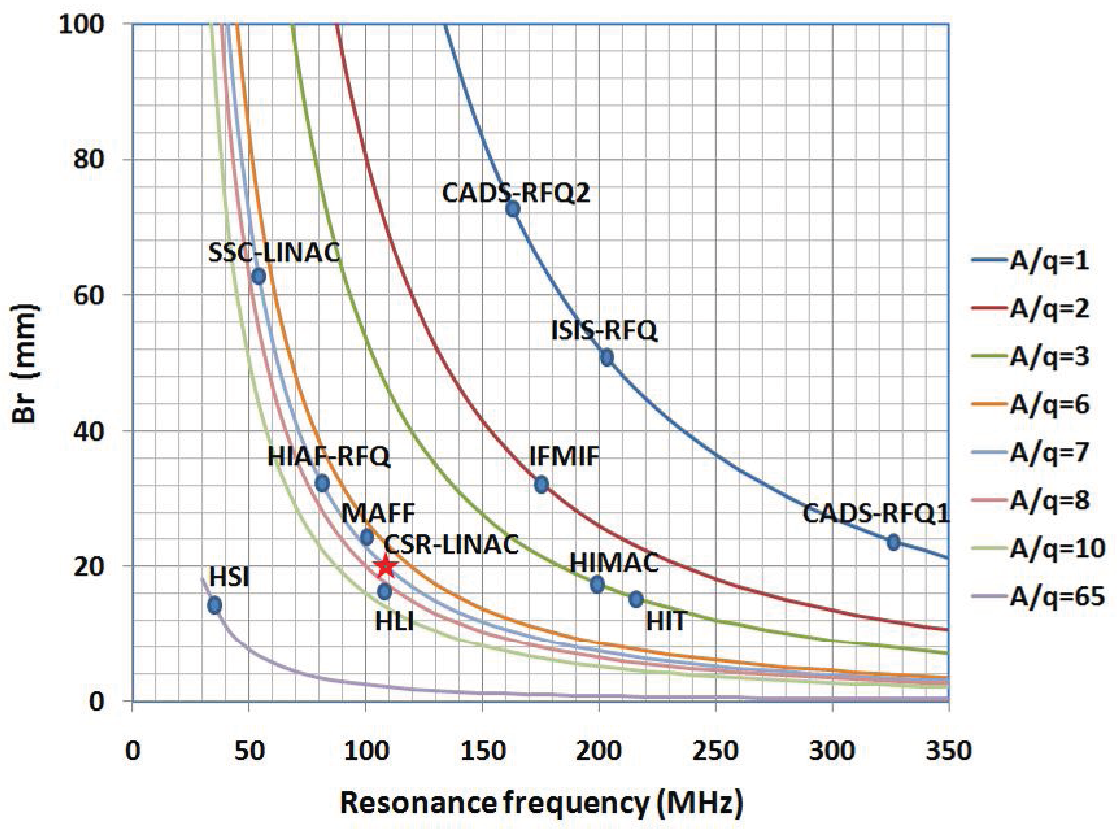}
\figcaption{\label{fig6} Br as the function of the frequency and A/q.}
\end{center}

Usually the focusing parameter $B > 4.5$, which supplies large enough transverse phase advance, is helpful for good beam transmission, and therefore 68.8KV is set as the vane voltage of RFQ through $U=\frac{E_sr}{\kappa}$. The RFQ input energy of 0.004MeV/u, which brings a fast bunch in RFQ, is reasonable and corresponding ECRIS extraction voltage is in a acceptable range of from 12KV to 28KV. The RFQ output energy of 0.3MeV/u, generally 2q/A MeV/u [7], can be well matched for the DTLs downstream. The main parameters of RFQ in CSR-LINAC are exhibited in the Table.2.

\begin{center}
\tabcaption{\label{tab1}  Main Parameters of the RFQ.}
\footnotesize
\begin{tabular*}{80mm}{c@{\extracolsep{\fill}}ccc}
\toprule
Ions & ${}^{12}\!C^{4+},{}^{86}\!Kr^{18+},{}^{129}\!Xe^{27+},{}^{209}\!Bi^{32+},{}^{238}\!U^{34+}$ \\ \hline
A/Q & 3 - 7 \\ \hline
Beam current & 3 emA  \\  \hline
$f_r$  & 108.48 MHz \\ \hline
$V_{vane}$ & 68.8 KV \\  \hline
Modulation  & 2.124 \\  \hline
$a_{min}$  & 2.623 mm \\  \hline
$L_{vane}$ & 3.07 m \\  \hline
$P_{dissipation}$  & 140 KW\\  \hline
$E_{in}/E_{out}$  & 0.004/0.3 MeV/u \\  \hline
Transmission & 95.3\% \\
\bottomrule
\end{tabular*}
\end{center}

To accomplish an efficient bunching process for the CSR-LINAC RFQ, an unconventional design approach, the so-called New Four-Section Procedure (NFSP) [8] has been employed. Abandoning the unreasonable constant-B law and the inefficient evolution manners of dynamics parameters adopted by the classic LANL four-Section Procedure (FSP) [9], the NFSP strategy deals with the accelerating beam for high transmission, small emittance growth and large error tolerance. Variable-B is adjusted to avoid the parametric resonance [10], especially the case of $\sigma_{0t}=\sigma_{0l}$ . The zero current phase advance is given by:
\begin{equation}
\sigma_{0t}^2 = \frac{B^2}{{8{\pi}^2}}+\Delta_{rf}
\end{equation}
\begin{equation}
\sigma_{0l}^2 = -\frac{Z{\pi}^2T{U_0}sin({\phi}_s)}{{AE_r{\beta}^2}}
\end{equation}
where B is the transverse focusing strength, the RF transverse defocusing factor $\Delta_{rf}=-\frac{1}{2}\sigma_{0l}^2$, T is the accelerating efficiency, $E_r$ is the rest mass per nuclei and $\phi_s$ is the synchronous phase, which directly influences the longitudinal capture efficiency. The beam non-resonant stable region requires $B>2\sqrt{3}\pi\sigma_{0l}$ and the larger B corresponds to larger error tolerance. However, the maximum surface electric field increases, up to the spark value, when B increases. In the gentle bunching (GB) section, B increased to balance the stronger growing transverse defocusing effect, so that the bunching speed can increase safely. When the acceleration (AC) section starts, the quickly increased beam velocity weakens the transverse defocusing effect, and B should accordingly fall down to avoid longitudinal emittance growth and to allow large bore aperture. Fig.7 exhibits RFQ parameters as a function of position z and the maximum surface electric field (23.2MV/m) appears at 80cm, where the longitudinal phase advance is closest to the transverse phase advance. The minimum aperture is 2.623mm at the AC section. Fig.8 shows the transverse beam envelope and the longitudinal phase space evolution along the RFQ. Beam loss occurs mainly at the end of GB section and the AC section. Fig.9 is the distribution in phase space. The FWHM energy spread at the exit of the RFQ is approximately 2.4\%.

\begin{center}
\includegraphics[width=8cm]{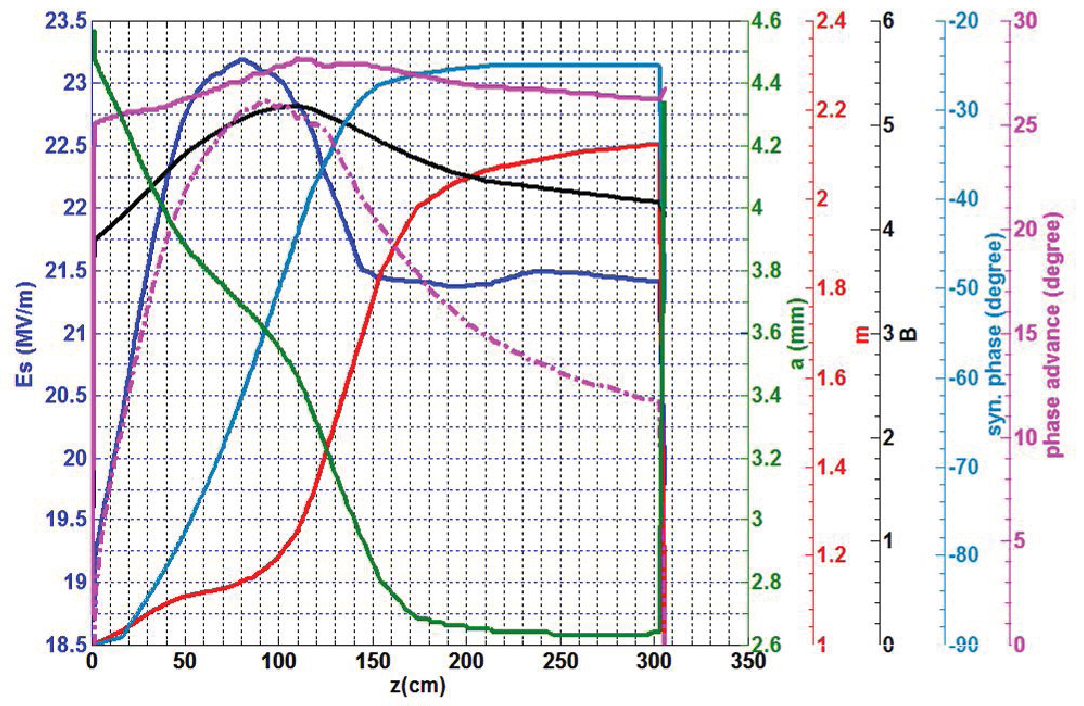}
\figcaption{\label{fig7} The RFQ beam dynamics parameters as a function of position z.}
\end{center}

\begin{center}
\includegraphics[width=7cm]{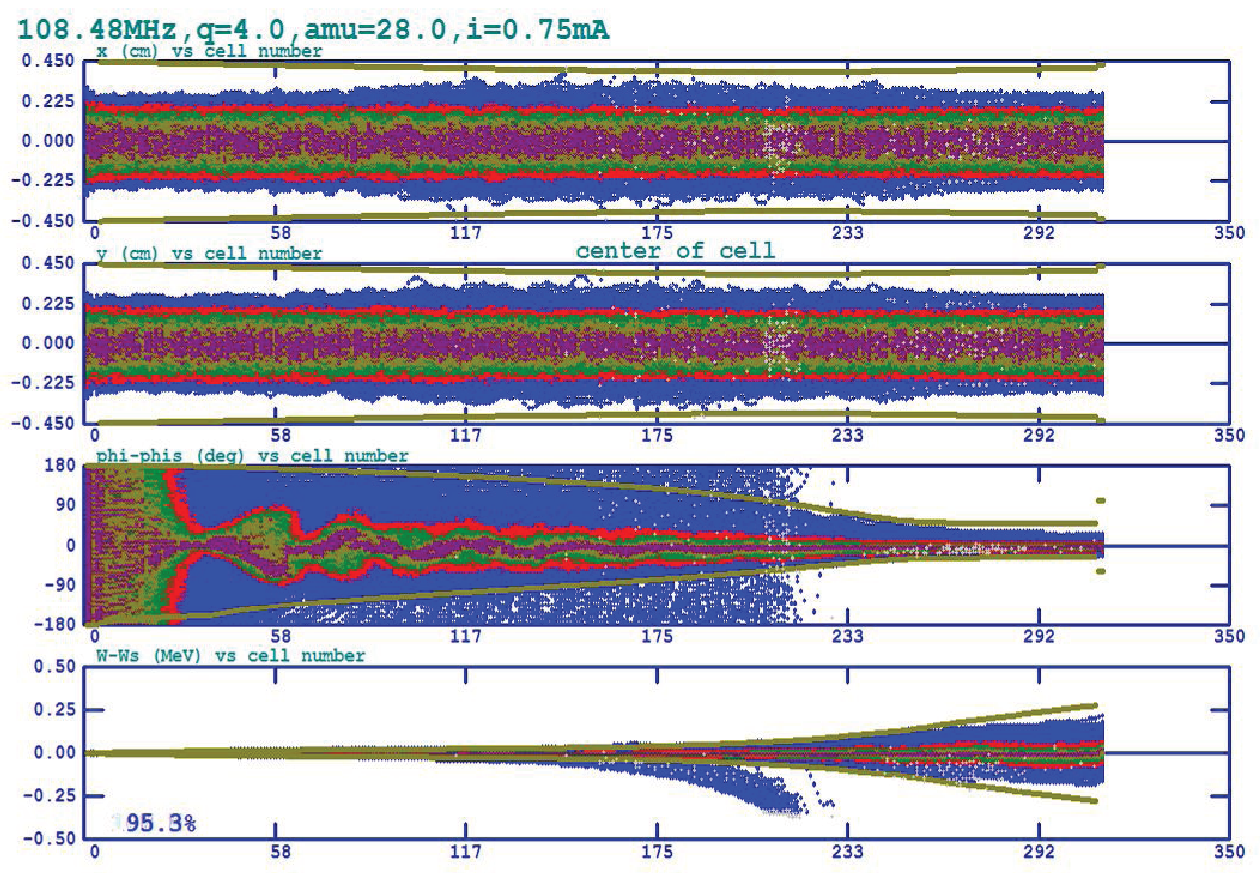}
\figcaption{\label{fig8} Beam transmission along the RFQ. Plots from top to bottom are the beam profiles in x and y planes, phase and energy spectrums respectively.}
\end{center}

\begin{center}
\includegraphics[width=7cm,height=5cm]{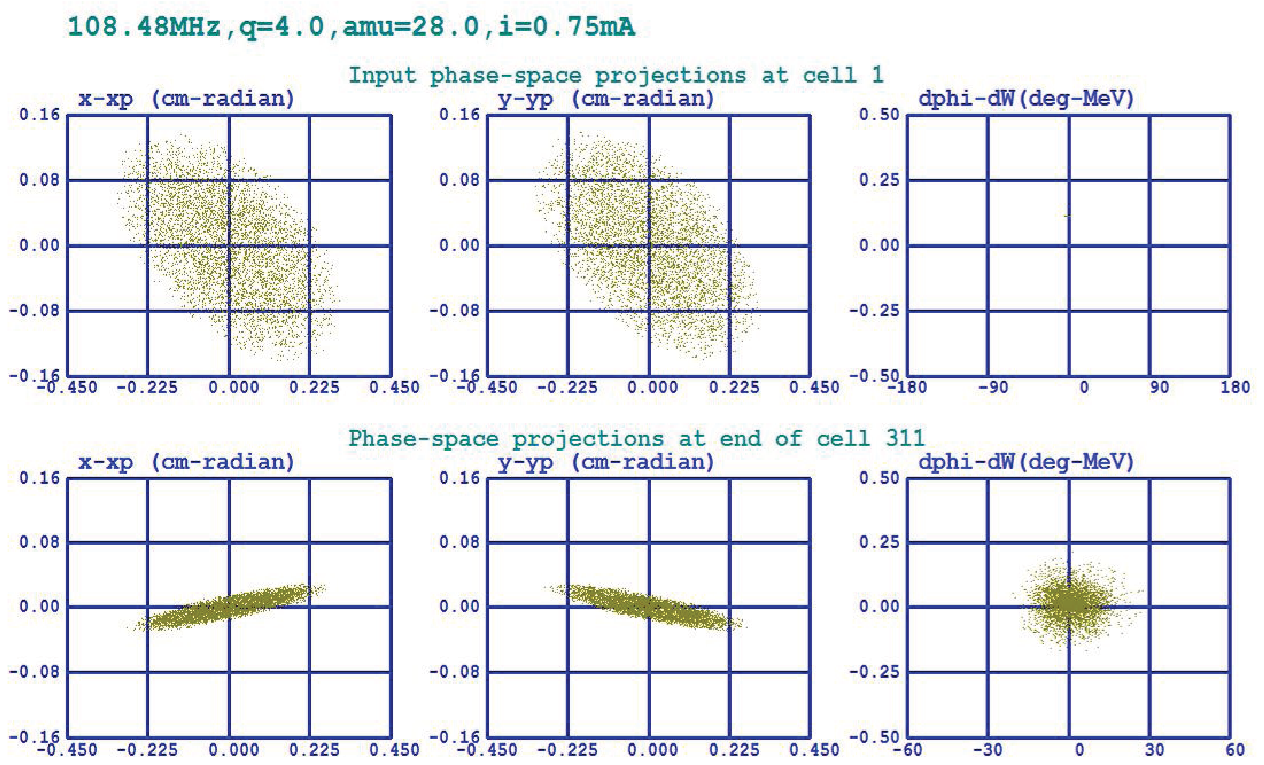}
\figcaption{\label{fig9} Transverse phase space projection at the entrance(upper) and exit (lower) of the RFQ.}
\end{center}

\section{DTL beam dynamics}
 The IH-DTL, which consists of 6 DTL tanks and a rebuncher, accelerates the ions from 0.3MeV/u to 7.272MeV/u at the resonance frequency of 108.48 and 216.96MHz. Its total effective accelerating voltage is 58.6MV and the total pulsed RF power is 2.8MW in the linac. The maximum accelerating gradient can reach 4.95MV/m at the last cavity with the resonance frequency of 216.96MHz. Table.3 exhibits the main parameters of the DTL section in CSR-LINAC.
\begin{center}
\tabcaption{\label{tab1}  Main Parameters of the DTL sections.}
\footnotesize
\begin{tabular*}{80mm}{c@{\extracolsep{\fill}}ccc}
\toprule
Tanks & 7 (including rebuncher)\\ \hline
A/Q & 3 - 7 \\ \hline
Beam current & 3 emA  \\  \hline
$f_r$  & 108.48/216.96 MHz \\ \hline
$L_{total}$ & 17.066 m  \\  \hline
$Gradient_{max}$ & 4.95 MV/m \\  \hline
$P_{rf}(total)$  & 2.8 MW\\  \hline
$E_{in}/E_{out}$  & 0.3/7.275 MeV/u \\  \hline
Transmission & $\ge$95\% \\
\bottomrule
\end{tabular*}
\end{center}

KONUS beam dynamics is applied to the drift tube linac (DTL) in CSR-LINAC [11], and IH (Interdigital H-mode) structure is adopted because of high shunt impedance [12]. The LORASR code is applied to the beam dynamics design of KONUS DTL. The period structure concept is proposed in the KONUS beam dynamics design. A KONUS period is composed of three sections with separated function. The first section consists of a few gaps with a negative synchronous phase of typically from -25¡ãto -35¡ãand acts as a rebuncher. Then the beam is injected into the main accelerating section with surplus energy and phase compared with a synchronous particle. Finally, the multi-gap section is followed by the transverse focusing elements, such as the magnetic quadrupole triplets [13]. In the beam dynamics design of the KONUS period structure, the key parameters to be chosen as following [14]: (1) effective accelerating voltage distribution, (2)cell number per section, (3)starting phase and energy in 0¡ã section.

For a given geometry, the relation of the peak electric field $E_p$ on the axis to the maximum surface electric field $E_{s}$ is illustrated by:
\begin{equation}
E_p = \kappa_{opt}(g,d)E_{s}
\end{equation}
where $\kappa_{opt}(g,d)$ depends on the accelerating gap length g, the tube diameter d and the tube pole geometry, which is independent to the resonance frequency when the cell size is much smaller than the RF wave length. The database of $\kappa_{opt}(g,d)$, simulated and calculated by CST Microwave Studio , shows that the corresponding relation of $E_{s}$ to $E_{p}$. The peak electric field limit $E_{p,limit}$ on the axis is obtained by:
\begin{equation}
E_{p,limit} = \kappa_{opt}(g,d)E_{spark}
\end{equation}
where $E_{spark}$ is the maximum surface electric field for the spark, which is 21.05MV/m for 108.48MHz and 27.38MV/m for 216.96MHz. The variation of $E_{p,limit}$ with gap length at different tube radius and resonance frequency, and the peak electric field distribution per cavity are present in Fig.10. As can be seen, the peak electric fields in the first DTL with tube radius of 10mm are lower than the corresponding peak electric field limit (black line) at the resonance frequency of 108.48MHz; the peak electric fields of the first 15 cells in the DTL2 with tube radius of 10mm are lower than the corresponding peak electric field limit (red line) at the resonance frequency of 216.96MHz; and the peak electric fields of the others downstream with tube radius of 11mm are lower than the corresponding peak electric field limit (green line) at the resonance frequency of 216.96MHz.

\begin{center}
\includegraphics[width=7cm]{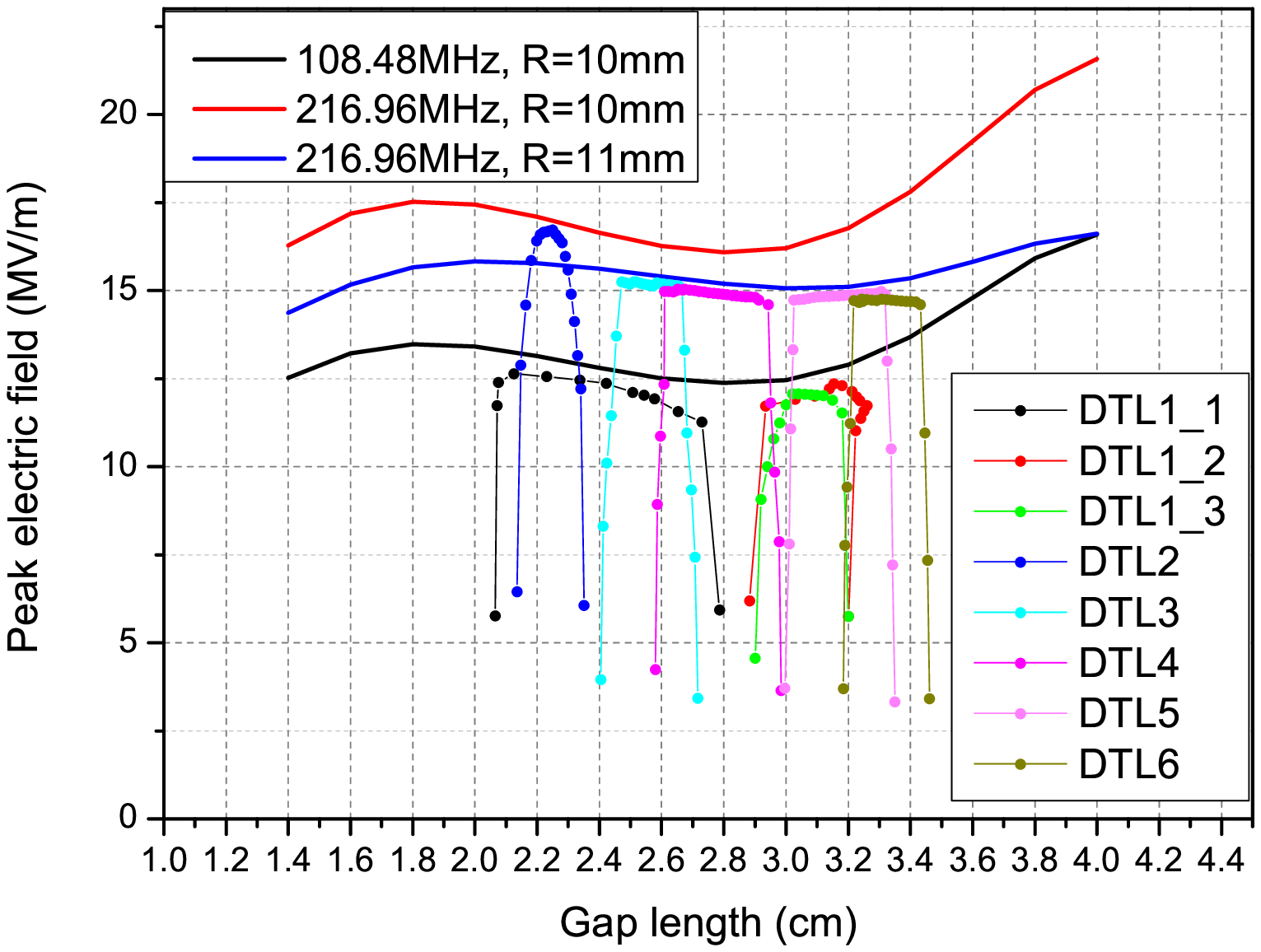}
\figcaption{\label{fig10} The dot-lines stand for the peak electric field distribution per cavity. The lines mean the peak electric field limit as the function of the gap length, at the different tube radius.}
\end{center}

In the LORASR code, the peak electric field ($E_p$ ) on the axis is calculated from the given effective accelerating voltage, and the reasonable effective voltage distribution is adjusted according to the peak electric field in Fig.10. Fig.11 shows the effective accelerating voltage distribution along the DTL section, and the maximum effective accelerating voltage is 0.561MV, which keeps at a safe region.
\begin{center}
\includegraphics[width=7cm]{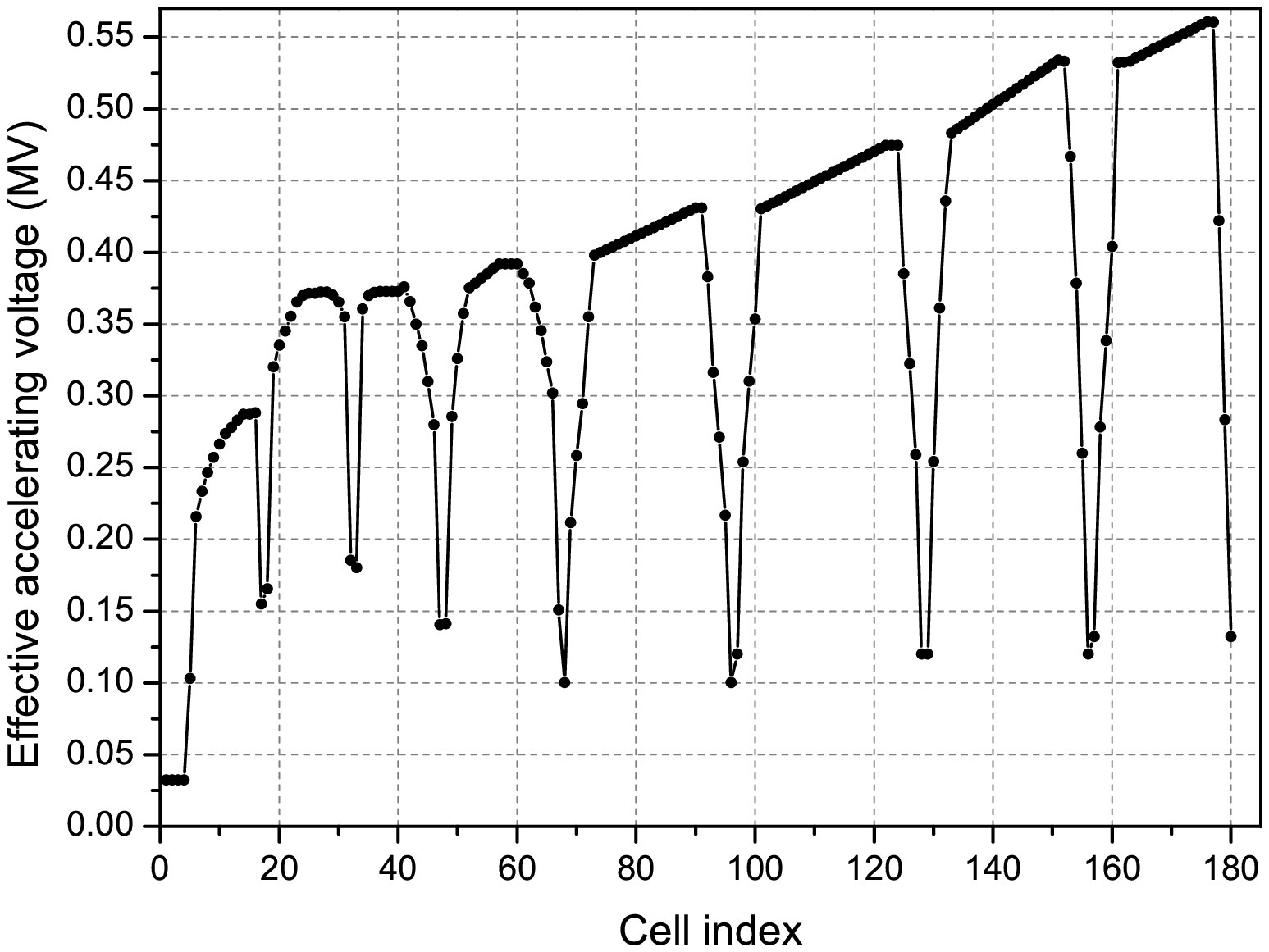}
\figcaption{\label{fig11} the effective accelerating voltage distribution per cell along the DTLs.}
\end{center}

Concerning the phase shift(for example at the transition $\phi_{sec1} \to \phi_{sec2}$) in the same cavity, the geometry length of the transition cell is adjusted by:
\begin{equation}
L_{shift} = (n+\frac{\phi_{sec2}-\phi_{sec1}}{180}){\frac{\beta\lambda}{2}}
\end{equation}
where n=1 at the transition to 0¡ãsection and n depends on the length of the quadrupole triplet where 0¡ã section transfers to negative phase section. The tank rf phases can be chosen independently when the transition gaps belong to different cavities. In the 0¡ãsection, the starting energy and phase are adjusted to get the desired beam parameters, which match the needs of the following sections. Generally, the final bunch centroid phase is in the range of -20¡ãto -30¡ã (seen in Fig.12), and also depends on the cell number in the 0¡ãsection.
\begin{center}
\includegraphics[width=7cm]{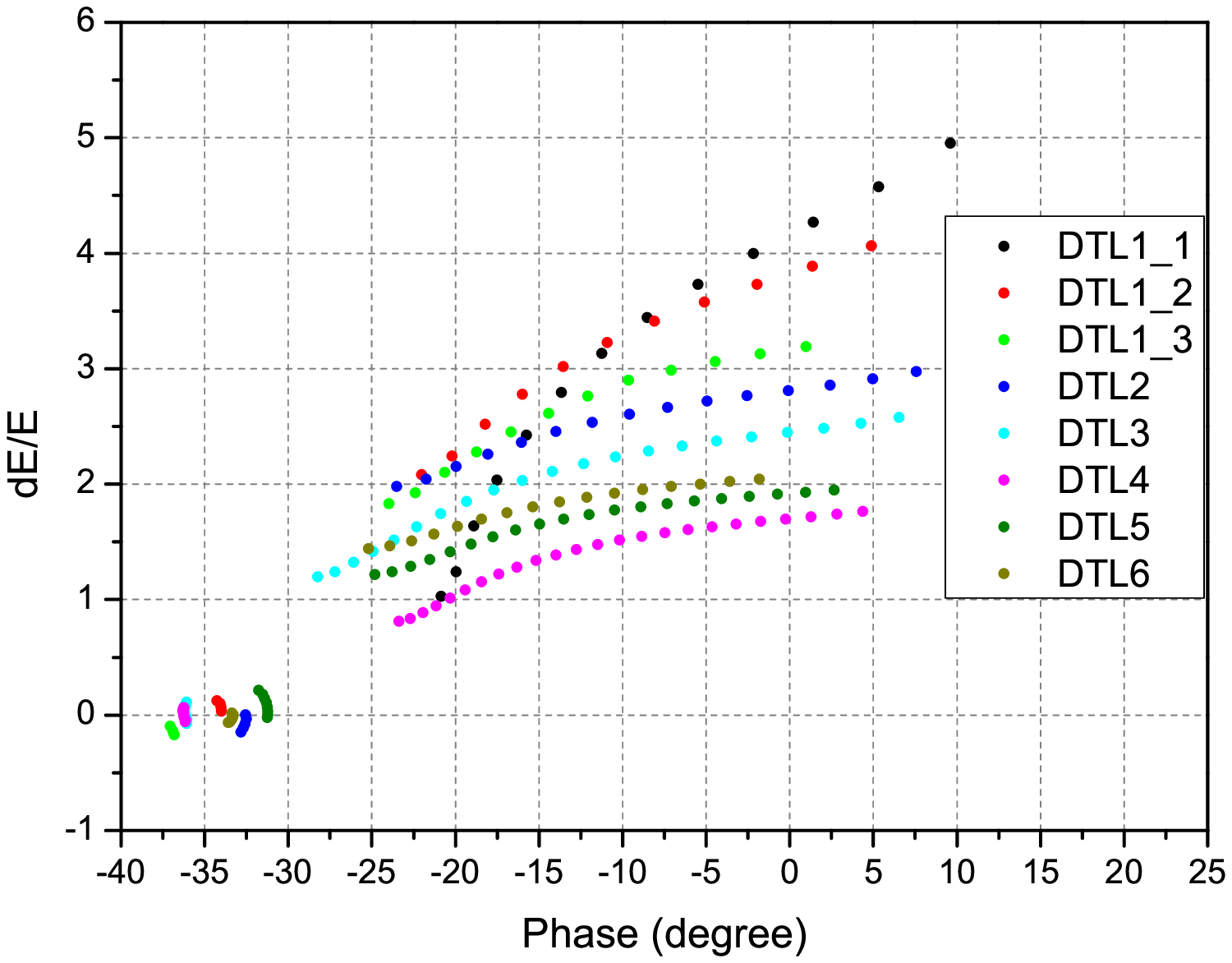}
\figcaption{\label{fig12} The evolution of the reference particle in the longitudinal phase space.}
\end{center}

Pole tip field up to $B_{max}=1.3T$ is available with conventional technology. At lower beam energy, the triplets must be installed within the resonator for the shortest possible drifts (QT3,QT4) between the sections, which make the mechanical design and RF tuning more complicated. With increasing beam energy, external lenses (QT5-QT10) are preferably used. A triplet section consists of four drift space and three magnetic quadrupoles, and parameters of magnetic quadruple triplets are listed in Table 4. The maximum quadrupole field gradient is 90T/m, corresponding to pole tip magnetic field of 1.17T for the triplet aperture diameter of 26mm, which keeps 10\% margin of the magnetic field limit,1.3T. Fig.13 shows transverse envelope evolution along position z. The beam envelope is smaller than 12mm in the drift tube section with aperture diameters of 20-22mm. Fig.14 exhibits the longitudinal relative energy spread and phase spread as a function of position z. The 95\% relative energy spread is smaller than 0.5\% at the end of DTL section, which can fill in the longitudinal acceptance of CSRm.

\begin{center}
\tabcaption{\label{tab4}  Quadruple Triplets parameters.}
\footnotesize
\begin{tabular*}{80mm}{c@{\extracolsep{\fill}}ccc}
\toprule
Triplets &  $L_{drift}$(mm) & $L_{eff,Q}$(mm) & Field(T/m)   \\ \hline
QT1 &  328/22/22/455.7 & 77/138/77 & 60/50/44   \\ \hline
QT2 &  127/22/22/142 & 77/138/77 & 58/51.5/57   \\ \hline
QT3 &  34.7/22/22/34.7 & 77/138/77 & 81/74.5/81   \\ \hline
QT4 &  35.5/22/22/35.5 & 77/138/77 & 80/78/78.5   \\ \hline
QT5 &  94/22/22/94 & 77/138/77 & 79/79.8/79   \\ \hline
QT6 &  94/22/22/94 & 92/162/92 & 79.1/82/82   \\ \hline
QT7 &  94/22/22/94 & 92/162/92 & 85/88.5/84.5   \\ \hline
QT8 &  109/22/22/109 & 92/162/92 & 80/85/80   \\ \hline
QT9 &  119/22/22/119 & 92/162/92 & 83/90/83   \\ \hline
QT10 &  119/22/22/- & 92/162/92 & 84/90/85   \\
\bottomrule
\end{tabular*}
\end{center}

\begin{center}
\includegraphics[width=7cm]{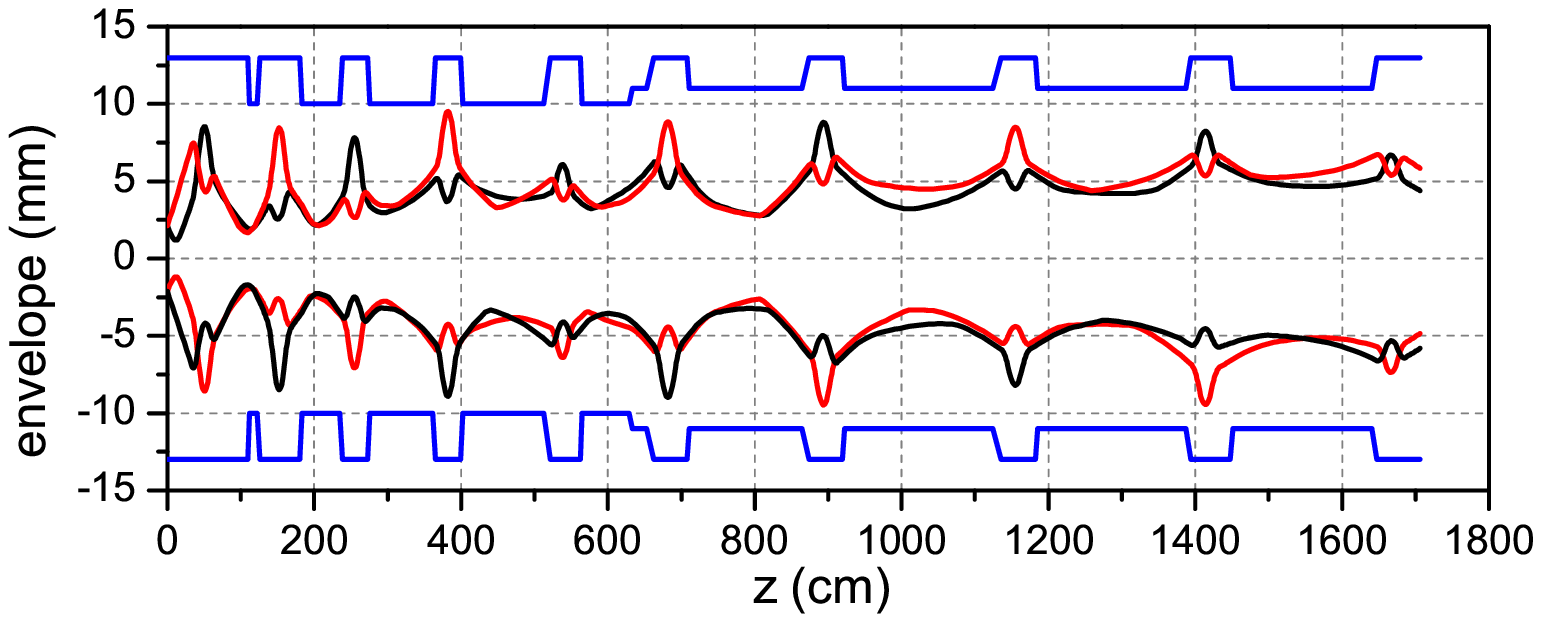}
\figcaption{\label{fig13} The x (red) and y (black) transverse envelope as a function of position z.}
\end{center}

\begin{center}
\includegraphics[width=7cm]{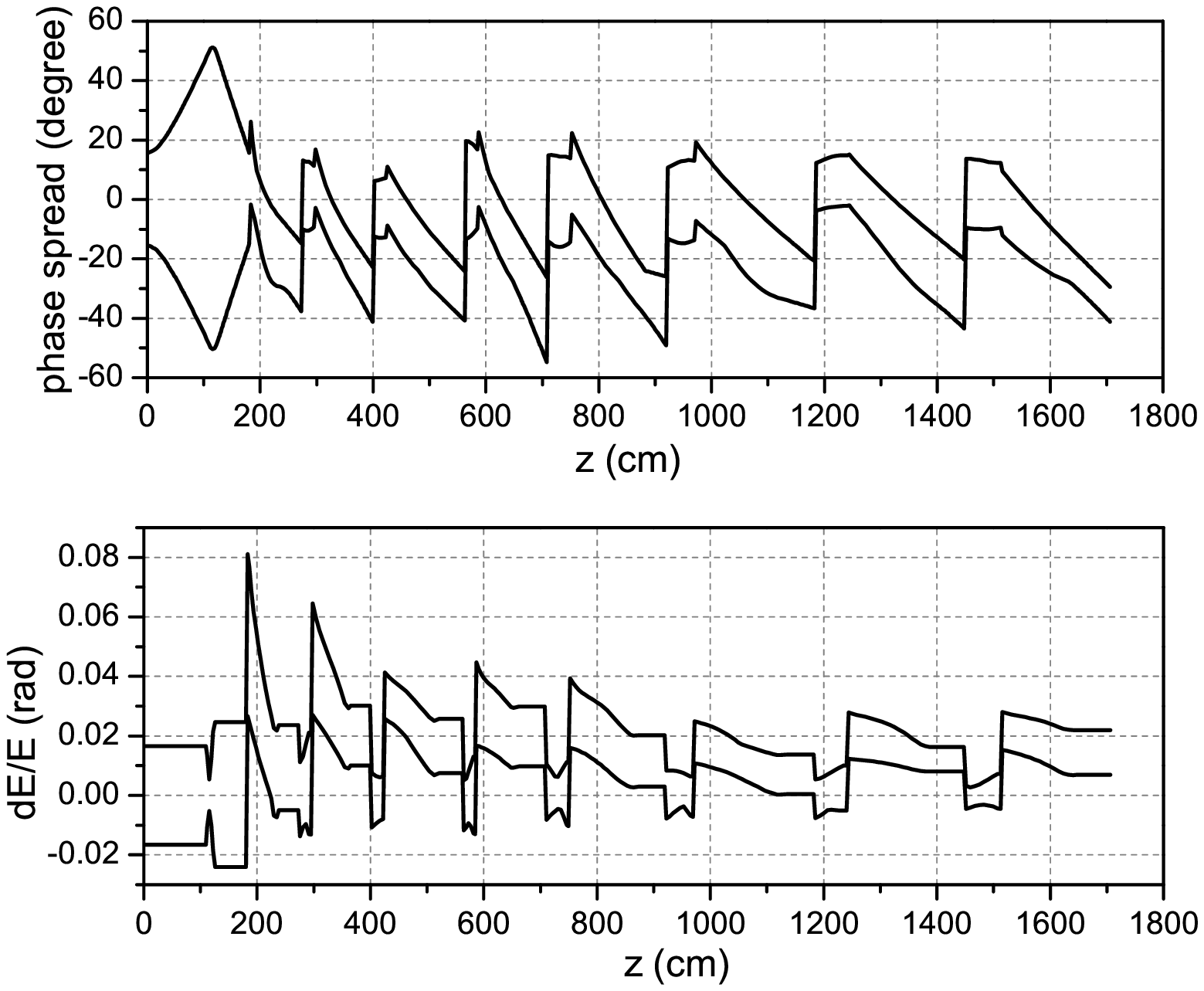}
\figcaption{\label{fig14} The longitudinal relative energy spread (upper) and phase spread (lower) as a function of position z.}
\end{center}

\section{Conclusion}
 The CSR-LINAC is proposed as the injector of CSRm, which is benefit to the capability improvement of HIRFL-CSR, and DIPM is achieved to double the operation time of the whole facility. The new linac injector can supply heavy ion with maximum mass to charge ratio of 7 and energy of 7.272MeV/u for CSRm. The beam current is as 10-100 times as that from SFC, which makes it possible to get higher charge state super-heavy ion by stripping and extract higher energy beam from CSRm.

 a new strategy, which is used for the determination of the main parameters (resonance frequency, RFQ vane voltage and DTL effective accelerating voltage) in heavy ion linac, is described in detail. 108.48MHz, as the basic resonance frequency of CSR-LINAC, is relatively high but brings high accelerating gradient for a compact injector. Taking advantage of the efficient NFSP strategy, a new beam dynamics design has been realized for CSR-LINAC RFQ. H-mode structure suited for the low and medium $\beta$ range, as well as KONUS beam dynamics concept for high gradient DTL, is applied to the main accelerating section in CSR-LINAC. The detail error study of CSR-LINAC will be present systematically in next paper. The CSR-LINAC project has been listed in the next 5-years plan in Institute of Modern Physics (IMP).

\section{Acknowledgements}
The authors want to give special thanks to Prof. Lu Yuanrong of Peking University for providing the LANL RFQ code, to Prof. U.Ratzinger for providing the LORASR code, and to the members of the CSR-LINAC work group for many discussions.
\end{multicols}

\vspace{-1mm}
\centerline{\rule{80mm}{0.1pt}}
\vspace{2mm}

\begin{multicols}{2}

\end{multicols}

\clearpage


\begin{thebibliography}{90}

\vspace{3mm}

\bibitem{lab1} Xia J W, Zhan W L, Wei B W, et al. The heavy ion cooler-storage-ring project (HIRFL-CSR) at Lanzhou[J]. Nuclear Instruments and Methods in Physics Research Section A: Accelerators, Spectrometers, Detectors and Associated Equipment, 2002, 488(1): 11-25.

\bibitem{lab2} Barth W. Commissioning of the 1.4 MeV/u high current heavy ion linac at GSI[J]. arXiv preprint physics/0008087, 2000.

\bibitem{lab3} Hutcheon D A, Bishop S, Buchmann L, et al. The DRAGON facility for nuclear astrophysics at TRIUMF-ISAC: design, construction and operation[J]. Nuclear Instruments and Methods in Physics Research Section A: Accelerators, Spectrometers, Detectors and Associated Equipment, 2003, 498(1): 190-210.

\bibitem{lab4} Yano Y. The RIKEN RI beam factory project: A status report[J]. Nuclear Instruments and Methods in Physics Research Section B: Beam Interactions with Materials and Atoms, 2007, 261(1): 1009-1013.

\bibitem{lab5} Wittkower A B, Betz H D. Equilibrium-charge-state distributions of energetic ions in gaseous and solid media[J]. Atomic Data and Nuclear Data Tables, 1973, 5(2): 113-166.

\bibitem{lab6} Wangler T P. RF Linear accelerators[M]. Wiley. com, 2008.

\bibitem{lab7} Staples J W. RFQs¡ªAn Introduction[C]. AIP Conference Proceedings. 1992, 249: 1483.

\bibitem{lab8} C. Zhang, A. Schempp, Nucl. Instr. and Meth. A 586 (2008) 153¨C159.

\bibitem{lab9} K.R. Crandall, R.H. Stokes, T.P. Wangler, RF quadrupole beam dynamics design studies, in: Proceedings of the 1979 Linac Accelerator Conference, Montauk,
NY, USA, September 9¨C14, 1979, pp. 205¨C216

\bibitem{lab10} Ilija M. Kapchinsky, Selected Topics in Ion Linac Theory[J], Lecture given at the University of Maryland, 1993.


\bibitem{lab11} Ratzinger U. The IH-structure and its capability to accelerate high current beams. Record of the 1991 IEEE PAC, San Francisco. 1991, 91: 3038-7.

\bibitem{lab12} Bres M, Chabert A, Foret F, et al. The interdigital H-type (IH) structure, an accelerating structure for low energy beams[J]. Particle accelerators, 1971, 2: 17.

\bibitem{lab13} H. Podlech, U. Ratzinger, A. Schempp, Institute of Applied Physics IAP, J.W. Goethe University Frankfurt/Main, Germany Design Study of A 3.7 AMeV Linac for A/Q Up To 8.5, 2009.

\bibitem{lab14} Tiede R, Ratzinger U, Podlech H, et al. KONUS beam dynamics designs using H-mode cavities[J]. Hadron Beam, 2008, 1: 2013.


\end{thebibliography}
\end{document}